\def\aap{{A\&A}}

\def\aj{{AJ}}
\def\araa{{ARA\&A}}
\def\apj{{ApJ}}
\def\apjl{{ApJ}}
\def\apjs{{ApJS}}

\def\mnras{{MNRAS}}

\def\pasp{{PASP}}
\def\bain{{BAIN}}

\documentclass[cup5b]{caps}
\usepackage{graphicx}
\usepackage{amssymb}
\usepackage{ociwsymp4}   

\HeadText{E. B. Jenkins}

\begin{document}

\pagenumbering{arabic}

\author[]{EDWARD B. JENKINS\\Princeton University Observatory}

\chapter{Interstellar Atomic Abundances}

\begin{abstract}

A broad array of interstellar absorption features that appear in the ultraviolet
spectra of bright sources allows us to measure the abundances and ionization
states of many important heavy elements that exist as free atoms in the
interstellar medium.  By comparing these abundances with reference values in the
Sun, we find that some elements have abundances relative to hydrogen that are
approximately consistent with their respective solar values, while others are
depleted by factors that range from a few up to around 1000.  These depletions are
caused by the atoms condensing into solid form onto dust grains.  Their strengths
are governed by the volatility of compounds that are produced, together with the
densities and velocities of the gas clouds.  We may characterize the depletion
trends in terms of a limited set of parameters; ones derived here are based
on measurements of 15 elements toward 144 stars with known values of $N$(H~{\sc
i})
and $N$(H$_2$).  In turn, these parameters may be applied to studies of the
production, destruction, and composition of the dust grains.  The interpretations
must be done with care, however, since in some cases deviations from the classical
assumptions about missing atoms in unseen ionization stages can create significant
errors.  Our experience with the disk of our Galaxy offers important lessons for
properly unravelling results for more distant systems, such as high- and
intermediate-velocity clouds in the Galactic halo, material in the Magellanic
Stream and Magellanic Clouds, and otherwise invisible gas systems at large
redshifts (detected by absorption features in quasar spectra).  Inferences about
the total (gas plus dust) abundances of such systems offer meaningful information
on their origins and/or chemical evolution.

\end{abstract}

\section{Introduction}\label{intro}

For those who study the interstellar medium (ISM), the manifold states that
atoms can assume in space are both a benefit and a hindrance.  On the
favorable side, the apportionments of atoms in various ionization stages
and the fractions that become bound in molecular compounds (including
solids) reveal the outcomes of fundamental processes that arise from
various physical and chemical influences, and these, in turn, disclose
the nature of nearby objects or conditions that created these
circumstances.  Conversely, if it turns out that we have a poor
understanding of the fractions of atoms that exist in unseen states, we
are hampered in our ability to gauge the relative total abundances of
different elements.  These two considerations underscore the importance
of our mastering the principles that govern how the atoms are subdivided
into different states within environments that range from the local part
of our Galaxy to the most distant gas systems that we can detect in the
Universe.  A major theme in this article will be to cover the underlying
principles of this topic.

Viewing atomic species by their absorption features in the spectra of
background continuum sources, such as stars or quasars, offers the
simplicity of not having to understand the details of how the atoms are
excited, as we must do if we are observing emission lines arising from
collisional excitation or recombination.  Moreover, the radiative
transfer of absorption features in the optical and ultraviolet regions
is straightforward, since the attenuation of light follows a simple
exponential absorption law.  Studies of absorption lines have revealed a
broad assortment of findings on the state and composition of
interstellar gases, and we review here many of these conclusions.

At this point, it is appropriate to digress and offer a brief historical
perspective.  The papers in this volume arose from one of four symposia
that celebrated the founding of the Carnegie Observatories exactly 100
years ago.  We may reflect on the fact that some two years after that
event, Hartmann  (1904) reported that stationary Ca~{\sc ii} lines in the
spectrum of the spectroscopic binary $\delta$~Ori arose from the
intervening medium, rather than from the atmospheres of the stars.  As
with many important discoveries, doubts about the interpretation
remained  (Young 1922), but eventually the interstellar nature of these
features was firmly established  (Plaskett \& Pearce 1933).  There soon
followed large surveys of interstellar lines, with Carnegie astronomers
playing an important role in the process  (Merrill et al. 1937; Adams
1949).   Nowadays, modern echelle spectrographs with very high resolving
powers show exquisite recordings of the complex superpositions of narrow
velocity features, and these data allow one to study either weak
features or those arising from very small parcels of gas along the line
of sight to the target stars  (e. g., Blades, Wynne-Jones, \& Wayte 1980; Welsh,
Vedder \& Vallerga 1990; Welty, Hobbs, \& Kulkarni 1994; Barlow et al. 1995;
Crane, Lambert \& Sheffer, 1995; Welty, Morton, \& Hobbs 1996; Crawford 2001;
Welty \& Hobbs 2001; Knauth, Federman, \& Lambert 2003; Welty, Hobbs, \& Morton
2003).

Against the backdrop of almost 100 years of studies of the interstellar
medium using absorption features in the visible part of the spectrum, we 
celebrate yet another anniversary.   Some 30 years ago the first papers
describing results from the {\it Copernicus\/} satellite  (Rogerson et
al. 1973a), launched in 1972, demonstrated the extraordinary utility of
using the ultraviolet part of the spectrum to study a vast array of
features arising from important interstellar atomic constituents 
(Morton et al. 1973; Rogerson et al. 1973b), along with our first
insights on the excitation and widespread presence of molecular hydrogen 
(Spitzer et al. 1973; Spitzer \& Cochran 1973).  Spitzer \& Jenkins 
(1975) reviewed many of the principal conclusions that came from this
initial venture into the ultraviolet.

Additional inferences on the nature of the ISM arose from the {\it
International Ultraviolet Explorer (IUE)\/}  (Kondo et al. 1987), which
bridged the {\it Copernicus\/} era and that of the {\it Hubble Space
Telescope (HST)\/}, launched in 1990.  {\it HST\/} still functions as a
premier facility for gathering useful observations of interstellar
absorption lines through the use of a new instrument, the Space
Telescope Imaging Spectrograph (STIS) that was installed in 1997 
(Kimble et al. 1998; Woodgate et al. 1998).  A comprehensive review of
the findings on interstellar abundances gathered from {\it HST\/} up to
1996 [using a first-generation spectrograph, the Goddard High-Resolution
Spectrograph (GHRS)], has been written by Savage \& Sembach  (1996). 
Finally, the {\it Far Ultraviolet Spectroscopic Explorer (FUSE)\/} 
(Moos et al. 2000) offers an opportunity to revisit the spectral region
$912-1185\,$ \AA\ with high sensitivity, a band that has been mostly
inaccessible since the days of {\it Copernicus}.

In parallel with investigations of the Galactic ISM using observatories
in Earth orbit, large-aperture telescopes on the ground have allowed us
to study absorptions by the same UV transitions arising from gaseous
material at very large redshifts.  Much of the progress in understanding
the abundances in these distant systems has drawn upon lessons learned
from studies of Galactic material, where we have better insights on the
fundamental abundance patterns (from the Sun and stars) and the physical
conditions.  Recent reviews by Pettini (2003) and Calura et al (2003)
summarize the current state of many of the conclusions on highly redshifted
systems that have arisen from such studies.

In the sections that follow, we will explore how the apparent abundances
of elements are altered by the effects of ionization and depletion onto
dust grains.  We bypass discussions of the techniques employed to study
the absorption features.  The reader is referred to a number of papers
that cover some key points on the methodology of this type of research 
(Cowie \& Songaila 1986; Jenkins 1986, 1996; Savage \& Sembach 1991;
Sembach \& Savage 1992), plus a list of possible pitfalls in its conduct 
(Jenkins 1987).

\section{Ionization Corrections}\label{ionization}

Of all the species that have absorption features in the visible part of
the spectrum, Ti~{\sc ii} is the only one that represents an element in its
favored stage of ionization within H~{\sc i} regions.  The remaining species,
Li~{\sc i}, Ca~{\sc i}, Ca~{\sc ii}, Na~{\sc i}, and K~{\sc i} all have an 
ionization potential less
than that of H~{\sc i}, with the consequence that they are mostly ionized to
higher stages by UV starlight photons that can easily penetrate the
regions.  One may attempt to derive the gas-phase abundance of an
element $X$ from solutions of the ionization equilibrium equation
involving the element's ionization rate $\Gamma(X)$ from starlight
photons, balanced against its rate of recombination governed by the rate
coefficients $\alpha(X,e)$ with free electrons and $\alpha (X,g)$ with
negatively charged or neutral, very small dust grains [at a rate that is
normalized to the hydrogen density $n$(H)\footnote{The rate coefficient
$\alpha (X,g)$ also depends on many incidental factors, such as the
character and concentration of dust grains, the local density of
starlight photons, and the local electron density.}]  (Weingartner \&
Draine 2001).  That is, in principle the balance
\begin{equation}\label{ioniz_equlib}
n(X^0)\Gamma (X) = n(X^+)[\alpha (X,e)n(e) + \alpha (X,g)n({\rm H})]
\end{equation}
may be solved to yield a total abundance of $X$ if the parameters are
well defined.  If the local values of $\Gamma(X)$ and the electron and
hydrogen densities $n(e)$ and $n$(H) are poorly known, it is still
possible to take Equation~\ref{ioniz_equlib} for element $X$ and divide it by
that for another element ($Y$) to derive a relative abundance of the two 
($X/Y$).  While this may seem to be an attractive solution for studying
abundance trends, recent comparisons made by Welty, Hobbs \& Morton 
(2003) indicate that different elements showing features in the
ultraviolet show mutually inconsistent values for the ratios of adjacent
ionization stages when Equation~\ref{ioniz_equlib} is invoked in the
respective cases.  These disparities may arise from either errors in the
atomic constants or, perhaps in some cases, from rapid changes in
conditions that prevent the equilibria from being fully established. 
Until these effects are better understood, comparing abundances from
minor ionization stages is probably a risky undertaking.

\begin{figure*}[t]
\includegraphics[width=1.00\columnwidth,angle=0,clip]{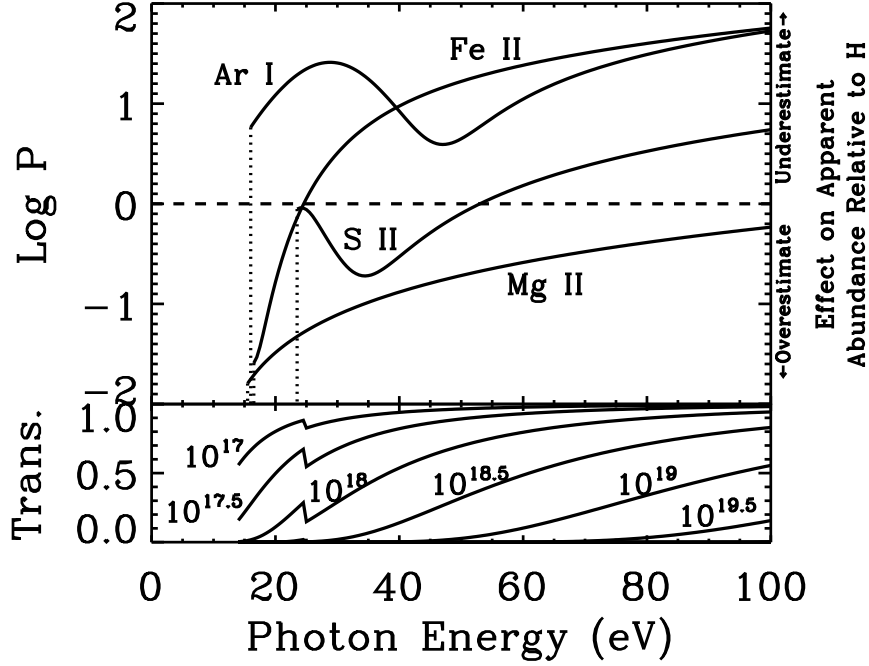}
\vskip 0pt \caption{
{\it Top section:} Relative ease of photoionizing Ar, S, Fe, and
Mg to ionization stages above the preferred ones (Ar~{\sc i}, S~{\sc ii}, 
Fe~{\sc ii}, and Mg~{\sc ii}) relative to that of partially ionizing H, 
expressed in the form of
the logarithm of $P$ defined in Eq.~\protect\ref{P} as a function of the
photon energy, under the simplified condition that the radiation field is
purely monoenergetic.  The amounts by which the derived abundances relative
relative to that of H are either underestimated ($\log P>0$) or overestimated
($\log P<0$) are defined by Eq.~\protect\ref{zeta}.  Curves for additional
elements are shown by Sofia \& Jenkins  (1998).  {\it Bottom section:} The
transmission of photoionizing radiation through different shielding column
densities of hydrogen, with the curves labeled according to different values
of $\log N({\rm H~I})$.}
\label{qplot_4}
\end{figure*}

For the major stages of ionization of different elements in H~{\sc i}
regions,
i.e., the lowest stages with ionization potentials greater than 13.6~eV,
it is generally assumed that nearly all of the atoms are in that stage. 
This is probably a trustworthy notion for individual regions with
hydrogen shielding depths $N({\rm H~I})>10^{19.5}\,{\rm cm}^{-2}$, but
below this column density photons with energies $<100\,$eV can penetrate
the region, as illustrated in the bottom section of Figure~\ref{qplot_4}. 
Once this happens, some fractions of both the element in question and
the accompanying hydrogen can be elevated to higher stages of
ionization, but by different amounts.  It follows that a comparison of
an element's most abundant stage to that of H~{\sc i} could lead to a
misleading abundance ratio, which differs from the true one by a
logarithmic difference
\begin{equation}\label{zeta}
\log\zeta(X)-\log\zeta({\rm H})=\log\left[{1+{n(e)\over n({\rm H})}\over
1+P(X) {n(e)\over n({\rm H})}}\right]~,
\end{equation}
where
\begin{equation}\label{P}
\zeta = ({\rm preferred~stage})/({\rm total})~~{\rm and}~~P(X) =
\left[{\Gamma(X)\alpha({\rm H})\over \Gamma({\rm H})\alpha(X)}\right]~.
\end{equation}
These equations are discussed in more detail by Sofia \& Jenkins 
(1998)\footnote{Sofia \& Jenkins refer to $\zeta$ as $\delta$, but $\delta$
is not adopted here because $\delta$ has often been used in much of the
literature to denote a depletion factor, as expressed here in
Eq.~\ref{depl_def}.}, who also present their more complex modified forms that
account for the additional effects of charge exchange with hydrogen. 
Alterations arising from charge exchange can be very important for certain
elements, such as N and O, which have large reaction rate coefficients 
(Field \& Steigman 1971; Butler \& Dalgarno 1979).

The curves for $\log P$ for several elements are shown in
Figure~\ref{qplot_4}.  This figure illustrates that under conditions of
partial ionization, $N$(Ar~{\sc i})/$N$(H~{\sc i}) underestimates the true
value of
Ar/H because Ar is more easily ionized than H, while the converse is
true for $N$(Mg~{\sc ii})/$N$(H~{\sc i}).  At first glance, S might appear to
be more immune to ionization corrections because its $\log P$ is not far
from zero.  However, the second ionization potential of S is 23.4~eV,
i.e., far above the 13.6~eV needed to ionize hydrogen.  Thus, there is
some danger that significant amounts of S~{\sc ii} could arise from a fully
ionized H~{\sc ii} region in front of a target star, and this might increase
the apparent abundance of S~{\sc ii} over that which should have been
identified with an H~{\sc i} region under study.  To illustrate how important
this effect can be, we note that for the extreme case of $\beta$~CMa the
ratio of foreground S~{\sc ii} to H~{\sc i} is 10 times solar  (Jenkins, Gry,
\& Dupin 2000).

The picture just presented is an oversimplification, but one that is
instructive.  In real life, the incident radiation has a distribution of
fluxes over different energies, and it is scattered and attenuated by
different amounts as deeper layers of a cloud are penetrated.  What we
view is the composite absorption produced by these layers.  One popular
approach for calculating these effects is to use the program CLOUDY 
(Ferland et al. 1998), which is designed to evaluate the various ion
fractions at different locations within a cloud that is irradiated from
the outside. This program has been made generally available to the
public (http://www.nublado.org) and is continuously updated. 
Another complication is that in some circumstances anomalously large
temperatures created by shock heating may lead to collisional
ionization, which can elevate even further the populations of 
high-ionization states  (Trapero et al. 1996).

Useful general discussions of the effects that can arise from a mixture
of ionized and neutral regions are given by Sembach et al.  (2000).  It
is clear that they are especially important in low-density gases within
about 100~pc of the Sun  (Jenkins et al. 2000; Jenkins, Gry, \& Dupin 2000; 
Gry \& Jenkins 2001).  Howk \& Sembach  (1999) point out the
dangers inherent in deriving abundances in damped Ly$\alpha$ systems,
and they back up their cautions by noting that the velocity profiles of
Al~{\sc iii} are very similar to those from species that are expected to
arise
from H~{\sc i} regions.  The
notion that such systems may have a considerable amount of partially
ionized gas is reinforced by the findings of Vladilo et al.  (2003)  who
noted that Ar~{\sc i} seems to be deficient relative to other
$\alpha$-capture
elements that are not expected to be appreciably depleted onto dust
grains (see \S\ref{depl} below).  By definition, damped Ly$\alpha$
systems have column densities $\log N({\rm H~I})>20.3$, well in excess
of the limit $\log N({\rm H~I})=19.5$ for shielding out photons with
$E\lesssim 100\,$eV.  Perhaps these systems are either (1) subjected to
appreciable fluxes with $E > 100\,$eV, (2) are made up of many thinner
regions that are each separately exposed to the ionizing radiation, or
(3) have strong sources of internal ionization arising from rapid star
formation.

\section{Depletions of Free Atoms onto Dust Grains}\label{depl}

\subsection{Basic Premise}

The fact that appreciable fractions of some elements in the ISM have
condensed into solid form is supported by a number of independent
observations.  First, starlight is absorbed and scattered by dust
particles or very large molecules---the opacity takes the form of
continuous attenuation in the visible and UV, together with discrete
absorption bands (2200 \AA\ feature, diffuse interstellar bands in the
visible, and infrared absorption features).  When compared with the
amount of hydrogen present, the magnitude of the attenuation indicates
that the proportions of some atoms in solid form are large (see the
review by Draine 2003).  Second, it is clear from
circumstellar emissions in the infrared that many types of evolved stars
cast off their dusty envelopes into the ISM.  While such dust probably
does not account for most of the material that is in solid form within
the general ISM, it probably provides nucleation sites for further growth in
dense clouds.  Finally, we may compare interstellar atomic abundances with
solar (or stellar) abundances in the Galactic disk and surmise that the
inferred depletions measure the fractional amounts of material hidden in some
kind of solid form (or within large molecules that are difficult to identify
spectroscopically).  Specifically, we may define a depletion of free atoms
using the usual bracket notation,
\begin{equation}\label{depl_def}
[X/{\rm H}]=\log(N(X)/N({\rm H}))-\log(X/{\rm H})_\odot
\end{equation}
and the amount of an element locked up in solids, relative to H, is
\begin{equation}\label{x_dust}
(X/{\rm H})_{\rm dust}=(X/{\rm H})_\odot (1-10^{[X/{\rm H}]})~,
\end{equation}
assuming that solar abundances are good reference values that truly
reflect the total abundances (see the second footnote in
\S\ref{reinvestigation}).  Much of the discussion that follows in this
paper concentrates on the findings on depletions, ones that are strikingly
evident from the UV absorption data, but which have also been evident even
from the visible absorption lines of Ti~{\sc ii}  (Wallerstein \& Goldsmith
1974; Stokes \& Hobbs 1976; Stokes 1978) and the features of Ca~{\sc ii} 
(Snow, Timothy, \& Seab 1983; Crinklaw, Federman, \& Joseph 1994)
(notwithstanding the inherent difficulties in dealing with the
uncertainties of the ionization equilibrium of the latter; see
Eq.~\ref{ioniz_equlib}).

\subsection{General Properties of the Depletion Trends}\label{trends}

To obtain a better understanding of the underlying causes of the
depletions of free atoms in space and the various factors that are
influential, it is important to identify systematic effects that occur
from one element to the next and from one environmental factor to the
next.  Three important effects have been observed: 
\begin{enumerate}
\item  Elements that reside within refractory compounds are more
severely depleted than ones that mostly condense into volatile ones. 
Field  (1974) was the first to identify this effect by comparing the
depletions of different elements toward $\zeta$~Oph with their
condensation temperatures, defined according to the temperature at which
some fixed proportion of an element in a gaseous cosmic mixture freezes
out into some solid compound(s).  The interpretation offered by Field
was that in cooling and expanding mass loss outflows from stars, the
dense, inner zones of the flows gave rise to nearly complete depletions
of elements making up the refractory compounds, whereas the outer, less
dense zones that were cool enough to freeze out more volatile substances
could not fully complete the process.  While this is an attractive
concept, we must also acknowledge that condensation temperatures are an
indication of general strengths of chemical bonds and may thus also
apply to the balance of creation and destruction of dust grains in the
general ISM  (Cardelli 1994).  A more modern presentation of the
correlations of depletions with condensation temperature (again for
$\zeta$~Oph) is shown in Figure~4 of the review article by Savage \&
Sembach  (1996).
\item  The strengths of depletions decrease for gas moving at large
velocities relative to the undisturbed material in a given region.  This
effect was first pointed out by Routly \& Spitzer  (1952) who compared
lines of Ca~{\sc ii} to those of Na~{\sc i}, and it has been substantiated in
greater detail by Siluk \& Silk  (1974), Vallerga et al.  (1993), and
Sembach \& Danks  (1994).  Misgivings that this effect could arise from
changes in ionization have been overcome by the ultraviolet data 
(Jenkins, Silk, \& Wallerstein 1976; Shull, York, \& Hobbs 1977), where
several ionization stages for certain elements could be observed.  The
interpretation of this effect is that gas at high velocities was
recently subjected to a shock that was strong enough to partly (or
completely) destroy the grains  (Barlow 1978a, b; Draine \& Salpeter
1979; Jones et al. 1994; Tielens et al. 1994).
\item  The strengths of depletions scale in proportion to the average
density of gas (i.e., $N$(H) divided by distance) along a line of sight 
(Savage \& Bohlin 1979; Harris, Gry, \& Bromage 1984; Spitzer 1985;
Jenkins, Savage, \& Spitzer 1986; Jenkins 1987; Crinklaw et al.
1994).  Such a measure is a very crude indication of conditions,
since there is no way to discriminate between very compact clouds with
high densities and a small filling factor from far more extended regions
having only moderately high densities but large filling factors. 
Nevertheless, the trend supports the propositions that the growth of
dust grains is markedly enhanced in dense regions and/or that within
these regions the destructive effects of shocks are muted.   Successful
alternative ways to gauge the predisposition for elements to deplete
include the fraction of hydrogen in molecular form  (Cardelli 1994) and,
toward very distant sources at high Galactic latitudes, simply $N$(H I) 
(Wakker \& Mathis 2000).
\end{enumerate}

\subsection{A Reinvestigation of the Depletion
Trends}\label{reinvestigation}

Seven years ago, Savage \& Sembach  (1996) summarized the interstellar
abundances derived from observations using the GHRS instrument aboard
{\it HST}.  Since that time, the continued success of {\it HST\/} (with
the STIS instrument that replaced GHRS) and emergence of {\it FUSE} have
enabled a significant number of important new surveys of interstellar
abundances to be carried out.  For this reason, it is useful to revisit
the general trends, now that a larger suite of results, both new and
old, can be compiled and analyzed.  Unlike previous studies that
directly compared the strengths of depletions of specific elements with
tangible properties of the lines of sight (e.g., Jenkins 1987 or Cardelli 
1994), we will initially ignore these external factors and instead
characterize depletions simply to arise as the result of some measure of each
element's general propensity to deplete, which works in combination with an
index $F_*$ for the overall strength of depletions (applied to all elements)
along the line of sight to a star.  This method has the advantage that links
in the depletions from element to element are evaluated directly.  As a
result, they are not weakened by the inevitable inaccuracies in the
relationships between individual depletions and some independently measured
quantity (either average density or the fraction of H$_2$).  After values of
$F_*$ have been defined for many different stars, we can then probe how they
correlate with the average density (or any other interesting attribute for a
line of sight).  When we investigate how depletions change under different
conditions, we enjoy the benefit of being able to draw upon a wide selection
of examples and not just those for a single element.

For the study presented here, abundances toward 144 stars were
identified from over 70 investigations reported in the literature (far
too numerous to cite here).  The targets were limited to those for which
we can extract reliable measures of hydrogen in the form of neutral
atoms and H$_2$, derived from measurements of the Ly$\alpha$ damping
profiles and the H$_2$ Lyman and Werner bands.\footnote{In some
instances, definite determinations for $N$(H$_2$) were not available. 
However, when upper limits for $N({\rm H}_2)$ were known to be smaller
than $0.02N$(H~{\sc i}), we could disregard the effect of H$_2$ in increasing
the value of $N({\rm H}_{\rm total})=N({\rm H~I})+2N({\rm H}_2)$.  Also,
based on general findings for H$_2$  (Savage et al. 1977; Rachford et
al. 2002), it is safe to ignore H$_2$ if the star's $B-V$ color excess
is less than 0.1, or (equivalently) $\log N({\rm H~I}) < 20.3$.}  Since
it is usually impossible (or very difficult) to identify the amounts of
hydrogen associated with individual velocity components along a line of
sight (without applying simplifying assumptions arising from
``undepleted'' species), only total column densities over all components
were considered.  Exceptions to this principle could be made for a few
cases where one component strongly dominated the total.

\begin{figure*}[t]
\includegraphics[width=0.48\columnwidth,angle=0]{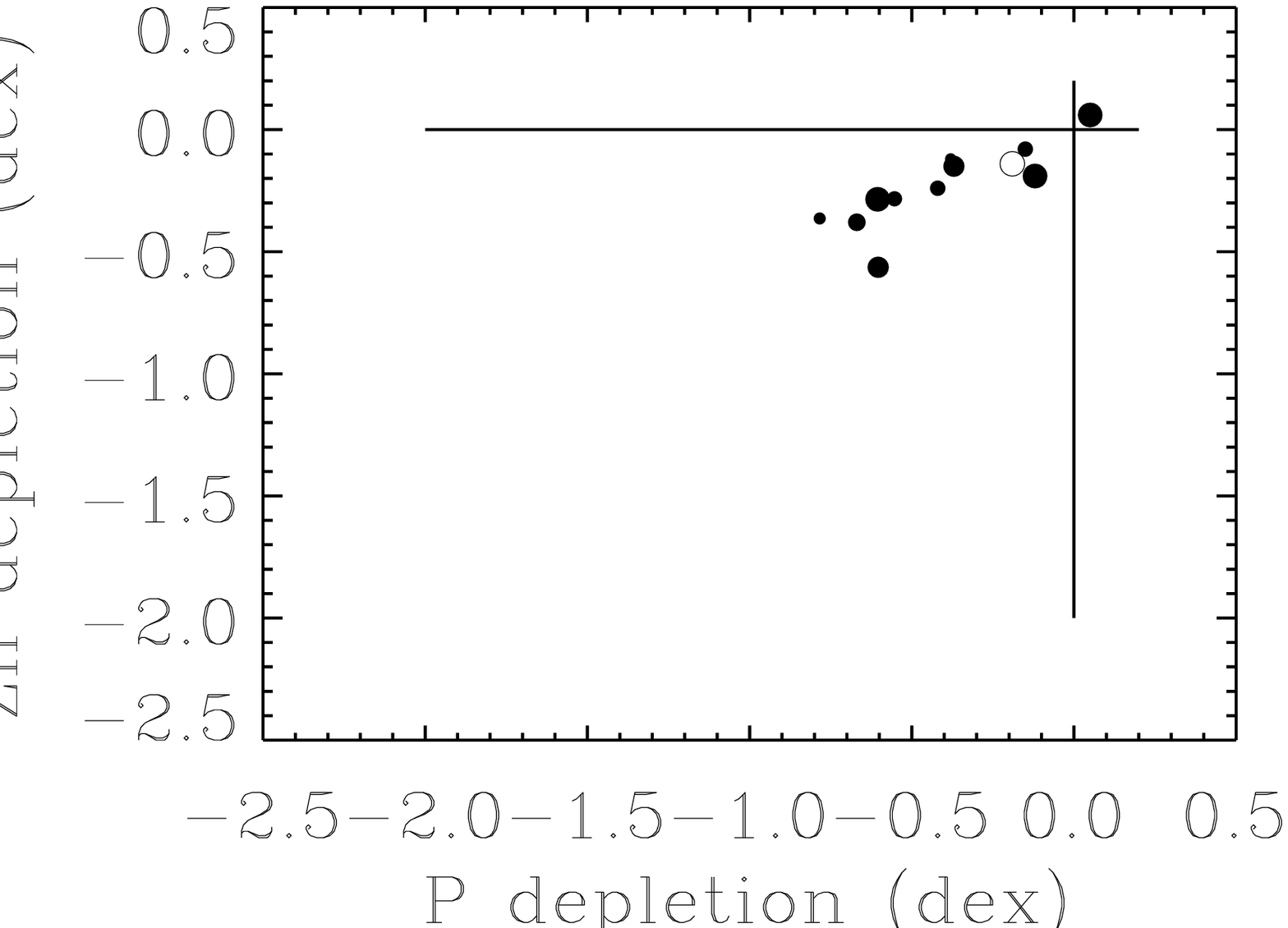}
\includegraphics[width=0.48\columnwidth,angle=0]{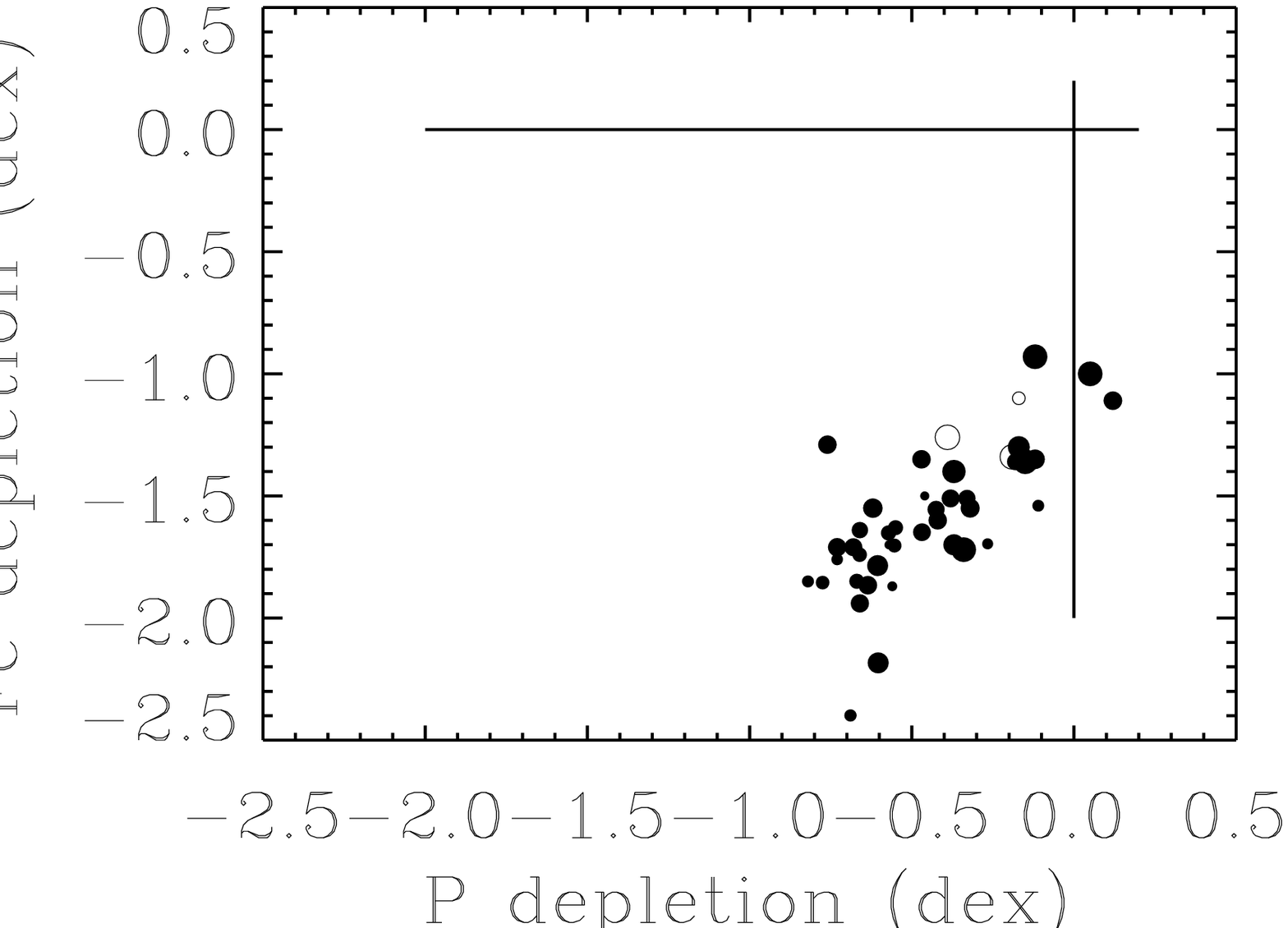}
\vskip 0pt \caption{
Two comparisons of depletions of pairs of elements along
different lines of sight: {\it Left panel:\/} Zn vs. $P$ and {\it Right
panel:\/} Fe vs. P.  Points with large diameters represent
determinations with small errors; open circles apply to cases where
$\log N({\rm H}) < 19.5$, where ionization effects may be important (see
\S\protect\ref{ionization}).}
\label{abund0}
\end{figure*}

We begin with the simplest assumption that from one line of sight to the
next, the depletions of different elements scale with each other in some
fixed proportion, as is illustrated for Zn and P shown in the left-hand
panel of Figure~\ref{abund0}.  However, this simple prediction does not
hold for all elements, as one can see for the case of Fe and P in the
right-hand panel of this figure.  This behavior seems likely to be explained
by the concept that typical dust grains have a mantle that is easily created
and destroyed in the ISM, and that this mantle surrounds a relatively
indestructible core that survives all but the most energetic (and very rare)
shock events  (Spitzer 1985; Jenkins, Savage, \& Spitzer 1986; Joseph 1988;
Spitzer \& Fitzpatrick 1993).  We are thus forced to consider at least three
parameters for a general law that characterizes the depletion of an element
$X$,
\begin{equation}\label{law}
[X/{\rm H}]=A_XF_*+A_{0,X},
\end{equation}
where $A_X$ represents the relative ease with which an element's
individual depletion changes as the general line-of-sight depletion
multiplier $F_*$ varies, and $A_{0,X}$ represents an offset attributable
to special compounds within the most rugged portions of the grains.  For
the study presented here, the normalization of $F_*$ was defined to make
a value of 1.0 equal to the depletions on line of sight toward
$\zeta$~Oph, a star that shows for one of its velocity components a
significant amount of depletion.  It is also a case that has probably
received the most attention of any star for depletions of a broad range
of different elements [see Table~5 of Savage \& Sembach  (1996) for a
summary].

\begin{table*}[t!]
\caption{Best-fit Depletion Constants}
\begin{tabular}{
c    
c    
c    
c    
c    
c    
c    
}
\hline \hline
{Elem.}&{Abund$_\odot$}&&&&& {Probability of}\\
{$X$} & {(H~=~12)} & {$A_X~(\pm 1\sigma$ error)} & {$A_{0,X}~(\pm
1\sigma$ error)} & {$\chi^2$} & {d.f.} & {a Worse Fit}\\
\hline

  C&8.39&$ -0.097\pm  0.208$&$ -0.148\pm  0.189$&  2.2&  6& 0.899\\
  N&7.93&$ -0.060\pm  0.075$&$ -0.080\pm  0.043$& 27.9& 15& 0.022\\
  O&8.69&$ -0.089\pm  0.067$&$ -0.050\pm  0.052$& 38.7& 28& 0.086\\
 Mg&7.54&$ -0.861\pm  0.046$&$ -0.248\pm  0.026$& 42.9& 62& 0.969\\
 Si&7.54&$ -1.076\pm  0.122$&$ -0.223\pm  0.033$&  5.0&  7& 0.660\\
  P&5.57&$ -0.967\pm  0.055$&$  0.088\pm  0.028$& 31.9& 52& 0.987\\
 Cl&5.27&$ -0.950\pm  0.108$&$  0.300\pm  0.068$& 41.3& 35& 0.215\\
 Ti&4.93&$ -2.226\pm  0.108$&$ -0.844\pm  0.064$& 30.5& 32& 0.543\\
 Cr&5.68&$ -1.373\pm  0.066$&$ -0.854\pm  0.036$& 16.8& 12& 0.158\\
 Mn&5.53&$ -0.685\pm  0.045$&$ -0.774\pm  0.025$& 52.7& 59& 0.704\\
 Fe&7.45&$ -1.198\pm  0.045$&$ -0.950\pm  0.023$& 59.7& 54& 0.275\\
 Ni&6.25&$ -1.440\pm  0.088$&$ -0.917\pm  0.046$& 31.0&  9& 0.000\\
 Cu&4.27&$ -0.408\pm  0.223$&$ -0.974\pm  0.204$&  7.7&  9& 0.561\\
 Zn&4.65&$ -0.522\pm  0.083$&$  0.042\pm  0.053$& 12.6& 12& 0.399\\
 Kr&3.23&$ -0.178\pm  0.134$&$ -0.116\pm  0.104$&  3.1&  9& 0.961\\
\hline \hline
\end{tabular}
\label{parameters}
\end{table*}

\begin{figure*}[t!]
{\centering \leavevmode
\includegraphics[width=\columnwidth]{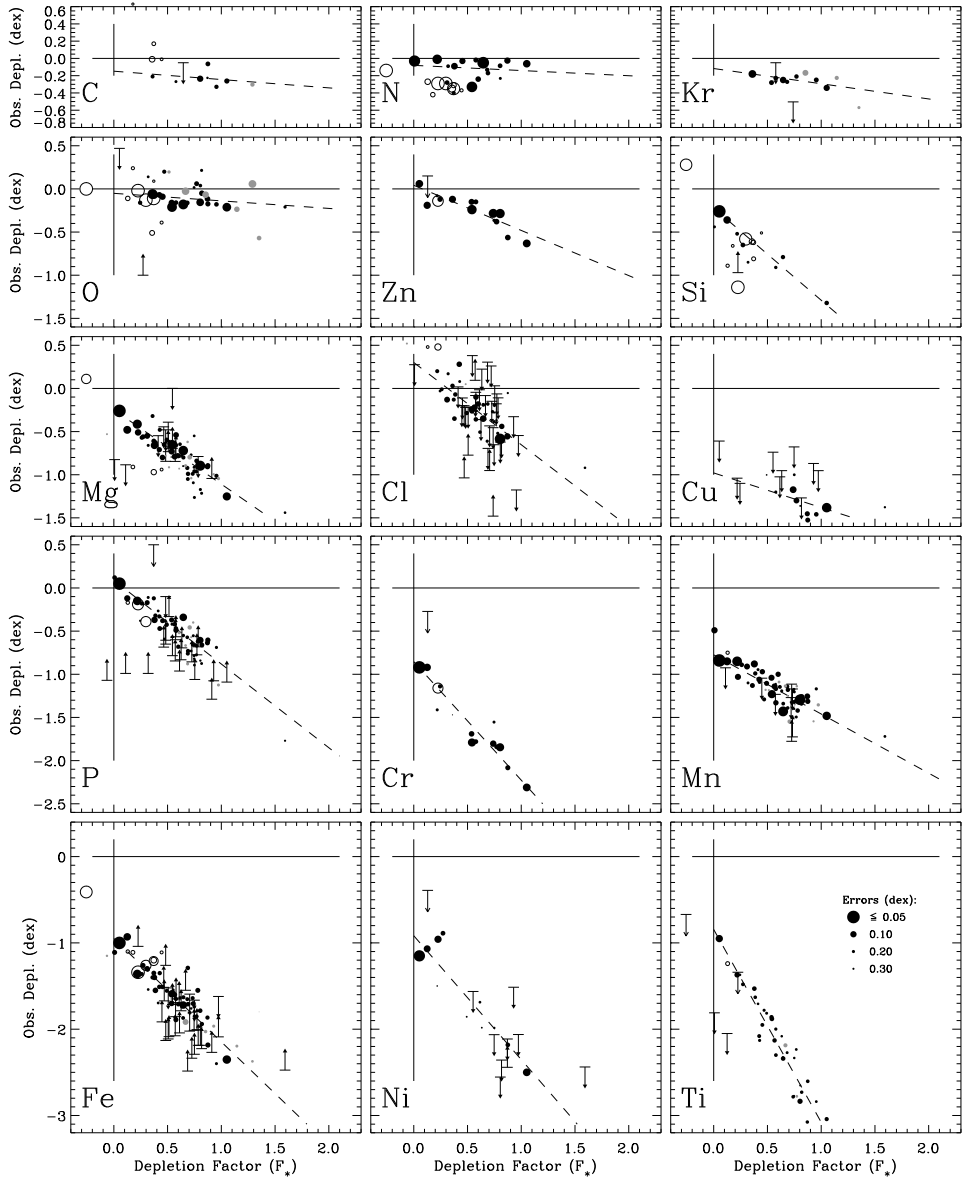}}
\vskip 0pt \caption{
Observed element depletions as a function of the line-of-sight
depletion multiplier $F_*$.  In each case, the dashed line represents
the linear fit defined by Eq.~\ref{law} with the constants $A_X$ and
$A_{0,X}$ listed in Table~\ref{parameters}.  Cases excluded in the
best-fit determinations of these constants include upper and lower
limits (arrows) and those with $\log N({\rm H}_{\rm total})<19.5$ (open
circles).  Errors in the observations are designated by the sizes of the
symbols (smaller symbols represent less certain results), according to
the legend in the lower right panel.  Cases for which fewer than three other
elements were available to establish a measure of $F_*$ are shown in
gray rather than black.  The open circles with negative $F_*$ correspond
to $\beta$~CMa, which has an extraordinarily large fraction of ionized
gas in front of it  (Gry, York, \& Vidal-Madjar 1985; Jenkins, Gry, \&
Dupin 2000).}
\label{abundf2}
\end{figure*}

Weighted least-squares determinations for the constants $A_X$ and
$A_{0,X}$ associated with all of the elements covered in this
study\footnote[2]{Initially, the elements S, Ar, and Ge were also
considered, but there were too few cases that met the acceptance
criteria to produce meaningful determinations of the depletion
parameters.  Reference abundances are shown in column~2 of
Table~\protect\ref{parameters} and were taken from the compilation by
Anders \& Grevesse  (1989), except for the revised solar abundances of C 
(Allende Prieto, Lambert, \& Asplund 2002), N, Mg, Si, Fe   (Holweger
2001), and O  (Allende Prieto, Lambert, \& Asplund 2001).  The revised
solar abundances appear to have resolved some outstanding discrepancies
that appeared to arise between the B-star abundances and the solar ones 
(Sofia \& Meyer 2001), which led to large ambiguities in $[X/{\rm H}]$
and $(X/{\rm H})_{\rm dust}$  (Snow \& Witt 1996).} are given in
Table~\ref{parameters}. Figure~\ref{abundf2} shows plots of the observed
depletions as a function of $F_*$; in each case a dashed line is drawn
to show the relationship expressed by Equation~\ref{law} for the appropriate
values of $A_X$ and $A_{0,X}$ shown in the table.  Within the
uncertainties of $A_X$, the elements C, N, O, and Kr all have slopes that
are consistent with zero, in accord with the findings of Cardelli et al. 
(1996), Meyer, Cardelli, \& Sofia  (1997), Meyer, Jura, \& Cardelli 
(1998) and Cartledge, Meyer \& Lauroesch (2003).  Table~\ref{parameters} also
indicates the values of $\chi^2$
for the fits, the associated degrees of freedom (d.f. equal to the
number of cases minus 2), and the probability of obtaining a worse value
of $\chi^2$ assuming that the errors are accurate and a linear fit of
the depletions $[X/{\rm H}]$ to $F_*$ is a correct model.  Values of
these probabilities that are unreasonably close to 0 or 1 probably
reflect unrealistically large or small errors assigned by the
investigators.  An exception is Cl, which seems to exhibit a genuinely
poor linear relationship.

\clearpage

\clearpage

\begin{figure*}[t]
\includegraphics[width=1.00\columnwidth,angle=0,clip]{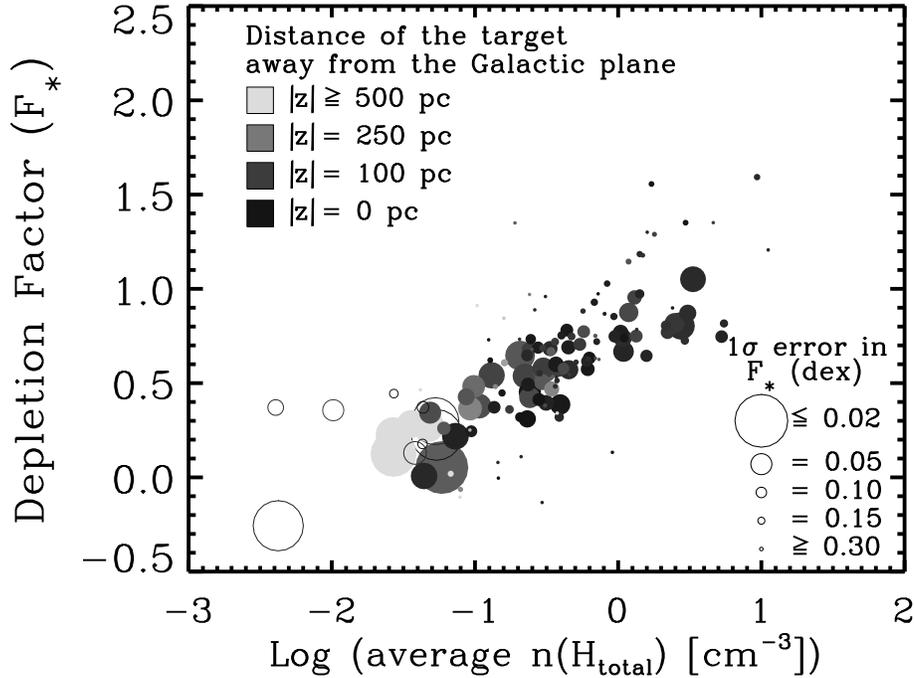}
\vskip 0pt \caption{
Relationship between the depletion factors $F_*$ toward
different stars and their respective average hydrogen densities along
the lines of sight.  Large symbols are more reliable than small ones, as
indicated by the legend in the lower right-hand corner, either because
more elements were sampled or the errors in the column densities were
smaller.  Open circles denote cases where ionization corrections might
be important because $\log N({\rm H~I})<19.5$.  The darknesses of the
filled symbols indicate the target star's distance above or below the
Galactic plane.}
\label{abundf1}
\end{figure*}

Figure~\ref{abundf1} shows how well the values of $F_*$ correlate with
the average hydrogen densities along the lines of sight.  There is a
clear trend between the two, which substantiates the statements made
earlier in \S\ref{trends} (point nr.~3).  The lack of any apparent
separation between the 
darker and lighter shaded points indicates that
for this sample, there is no clear effect that distinguishes paths
toward stars in the lower halo from those well within the Galactic plane
(but see \S\ref{halo}).

Savage \& Sembach  (1996) summarized depletions of the elements Mg, Si,
S, Mn, Cr, Fe, and Ni in terms of broad categories that were labeled 
``warm disk'' and ``cool disk,'' in addition to identifying
circumstances that included some gas arising from the Galactic halo (see
their Table~6).  As a point of reference, if one takes their listed
depletions and evaluates the corresponding values of $F_*$ for all of
the elements\footnote{For Mg, one should raise the Mg abundances listed
by Savage \& Sembach by 0.29~dex, in recognition of the improved
$f$-values of the 1240 \AA\ doublet determined by Fitzpatrick  (1997).
Likewise, abundances of Ni should be increased by 0.27~dex (Fedchak \& Lawler
1999)}, the warm disk gas is equivalent to $F_*=0.21$ while the cool disk
material is equivalent to $F_*=0.99$.  The rms dispersion of these two
numbers from one element to the next is 0.05.  From strong deficiencies of
Al~III compared to S~III along certain lines of sight, Howk \& Savage (1999)
concluded that dust grains reside within ionized regions as well.

\section{Insights on the Composition of Interstellar Dust}\label{dust}

\begin{figure*}[t] 
\includegraphics[width=1.00\columnwidth,angle=0,clip]{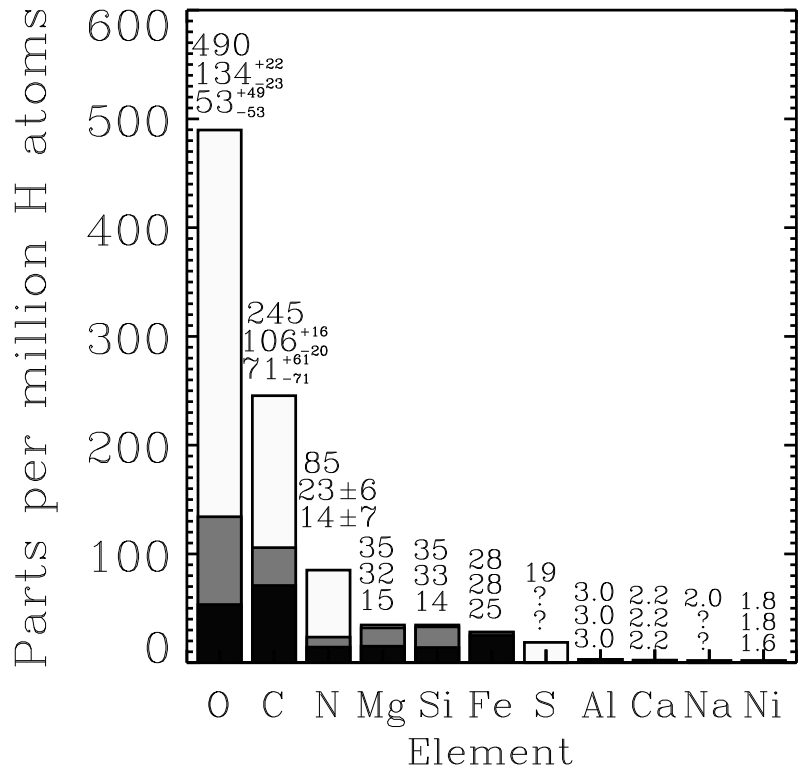}
\vskip 0pt \caption{
Depictions of elements having solar abundances greater than
$10^{-6}$ times that of H (excluding H and the noble gases He and Ar),
with the horizontal positions arranged according to the ranks in total
abundances shown by the heights of the bars.  The interior of each bar
is broken down by the element's fraction in gaseous form (upper, clear
portion) and solid form $(X/{\rm H})_{\rm dust}$ evaluated from
Eqs.~\protect\ref{x_dust} and \protect\ref{law} according to whether
$F_*=1.0$ (gray plus black lower portions) or $F_*=0.0$ (only the black
portion).  Numbers over each bar state the values (and errors, if more
than 3~ppm) corresponding to the tops of the black, gray and clear zones.}
\label{abundf5}
\end{figure*}

As an aid to help us visualize the composition of dust grains,
Figure~\ref{abundf5} indicates what fractions of the more abundant
elements exist in the simplest characterizations of (1) resilient grain
cores depicted by the lower, black portions of the bars, (2) the outer,
somewhat interchangeable forms (sometimes in mantles and other times as
free atoms) shown by the gray extensions, and (3) atoms perpetually in
free gaseous form shown by the clear portions that extend to the top of
each bar (with a length equal to the total abundance).   The divisions
between the three zones are based on evaluations of Equation~\ref{law} for
$F_*=1.0$ (tops of the gray zones) and  $F_*=0.0$ (the boundary between
the gray and black zones).  The presumption here is that these two
values of $F_*$ offer good representations of the quantities of material
found in grains that are, respectively, either stripped or well
nourished.  Note from a previous statement in \S\ref{reinvestigation}
that it is not clear from the observations that C, N, and O show
significant changes in depletion as $F_*$ progresses from one extreme to
the other, so in truth the differences between the gray and black zones
are not well established.  Two elements shown in the figure were not
studied here, primarily because there were problems with the ionization
effects discussed in \S\ref{ionization}: Na is seen only as Na~{\sc i} (an
unfavored stage of ionization) in the visible, and significant
contributions of S~{\sc ii} may arise from H~{\sc ii} regions.  Sofia et al. 
(1994), Spitzer \& Fitzpatrick  (1993), and Fitzpatrick \& Spitzer  (1994) 
have considered the possible mix of compounds in grains that were consistent
with the observed depletions; the current state of this topic is reviewed by
Draine  (2003).

\section{Outside the Galactic Plane}\label{outside}
\subsection{Lower Halo}\label{halo}

In order to study the composition of gases in the lower halo, it is
necessary to eliminate the inevitable foreground contributions arising
from material in the Galactic disk.  This can be accomplished by
selecting velocity components that are displaced from that of the local
gas by virtue of either differential Galactic rotation or special kinematics
(usually infall, as with material that is often referred to as ``intermediate
velocity clouds''\footnote{Not to be confused with the high velocity clouds
discussed in \S\protect\ref{hivel}.}).  However, in this circumstance it is
difficult to obtain useful information on the amount of hydrogen that
accompanies this gas.  That is, $N$(H~{\sc i}) derived from the Ly$\alpha$
damping wings applies to all of the hydrogen present, regardless of velocity. 
Measures of 21-cm emission show hydrogen as function of velocity, but the
results are often not very satisfactory because the radio beam covers a solid
angle in the sky that is much larger than a typical cloud size and thus may
detect material that is not on top of the source viewed in the UV.  Also,
some of the signal may arise from hydrogen beyond the target if it is a halo
star and not a distant QSO or AGN.

One method of overcoming the lack of good information on hydrogen is to
use some other element as a proxy, preferably one that is known to be
very lightly depleted.  For instance, Sembach \& Savage  (1996)
summarized the abundances of various elements relative to the
corresponding amounts of Zn in halo gas components along various
sightlines.  Those findings are reproduced in Table~\ref{halo_abund} in
the form of a value followed by the observed dispersion of results from
one halo star to the next.  For comparison, the table also shows similar
ratios for disk gas at two extremes for the depletion index, $F_*=0.0$
and 1.0.

\begin{table*}[t]
\caption{Abundances Compared to Zn for Halo and Disk Gas}
\begin{tabular}{
c    
c
c
c
c    
c
c
c    
c
c
c    
c
c
c    
}
\hline \hline
&&&&Halo&&&rms&&&Disk with&&&Disk with\\
Ratio&&&&Value&&&Dispersion&&&$F_*=0.0$&&&$F_*=1.0$\\
\hline
{[Mg/Zn]}\footnotemark[1]&&&&$-0.23$&&&$<0.19$       &&&$-0.29$&&&$-0.63$\\
{[Si/Zn]}                &&&&$-0.26$&&&\ \ \ \ \ 0.14&&&$-0.27$&&&$-0.82$\\
{[S/Zn]}                 &&&&$-0.05$&&&\ \ \ \ \ 0.14&&&\dots  &&&\dots  \\
{[Cr/Zn]}                &&&&$-0.51$&&&\ \ \ \ \ 0.13&&&$-0.90$&&&$-1.75$\\
{[Mn/Zn]}                &&&&$-0.61$&&&\ \ \ \ \ 0.09&&&$-0.82$&&&$-0.98$\\
{[Fe/Zn]}                &&&&$-0.64$&&&\ \ \ \ \ 0.04&&&$-0.99$&&&$-1.67$\\
{[Ni/Zn]}\footnotemark[1]&&&&$-0.56$&&&\ \ \ \ \ 0.07&&&$-0.96$&&&$-1.88$\\
\hline
\end{tabular}
\label{halo_abund}
\footnotetext[0]{$^*$As noted in an earlier footnote, the originally
quoted abundances of Mg and Ni have been increased by 0.29 and 0.27~dex,
respectively.}
\end{table*}

We must be cautious when interpreting the data shown in
Table~\ref{halo_abund}.  The column densities of Zn~{\sc ii} typically
indicate that $\log N({\rm H~I})\approx 19.5$, which means that some of the
gas could be partially photoionized, and this in turn would indicate
that the column density comparisons may be subject to the ionization
corrections discussed in \S\ref{ionization}.  Nevertheless, even with
the possibility that the apparent abundances may be perturbed in this
manner, we can still formulate some useful conclusions.  For instance,
we note from Figure~\ref{qplot_4} that Fe~{\sc ii} is more easily ionized to
higher stages than Mg~{\sc ii}, regardless of the characteristic photon
energy.  However, according to the entries in Table~\ref{halo_abund},
${\rm [Fe/Zn]}_{\rm halo}>{\rm [Fe/Zn]}_{\rm disk}$, while ${\rm
[Mg/Zn]}_{\rm halo} \approx {\rm [Mg/Zn]}_{\rm disk}$ for $F_*=0.0$.  It
is therefore safe to say that Fe is liberated from dust grains more
easily in the halo than in the disk, regardless of the strength of the
ionization (or the value of $F_*$ adopted for the disk gas).  Evidently,
in the halo the amount of residual iron left in grain cores is not as
large as in the most lightly depleted gas in the disk, i.e., ${\rm
[Fe/H]}_{\rm halo} > A_{0,Fe}=-0.95$.  This effect might arise from the
condition that much of the gas in the halo was heavily shocked as it was
ejected from the disk and thus the grains were more completely
destroyed, or, alternatively, that perhaps ejecta from Type~Ia
supernovae in the halo have had a chance to enrich the Fe in a time that
is shorter than the circulation time between the disk and halo  (Jenkins
\& Wallerstein 1996).

\subsection{High-velocity Clouds}\label{hivel}

We now move on to examine the abundances in distinct cloud complexes at
high Galactic latitudes that are moving at extraordinarily large
velocities  (Oort 1966; Mathewson, Cleary, \& Murray 1974), i.e., with
kinematics clearly not associated with differential Galactic rotation. 
A current and very comprehensive review of this material has been
presented by Wakker  (2001).  Here, our objective is not to assume some
reference abundances and then study the condensation into dust grains,
but instead, we seek to use abundances as a tool to gain a better
understanding about the origin and history of the material.  However, we
must still apply the concepts that we have considered earlier and
account for the alterations caused by ionization effects and the removal
of atoms as they condense into dust grains.

\begin{table*}[t]
\caption{Observed Abundance Ratios in the Magellanic Stream (MS) and
Complex C (C)}
{\footnotesize
\begin{tabular}{
c    
c    
c    
c    
c    
c    
c    
c    
c    
}
\hline \hline
&{[N~{\sc i}/}&{[O~{\sc i}/}&{[S~{\sc ii}/}&{[Si~{\sc ii}/}&{[Fe~{\sc
ii}/}&{[Ar~{\sc i}/}&$\log$\\
Object&{H~{\sc i}]}&{H~{\sc i}]}&{H~{\sc i}]}&{H~{\sc i}]}&{H~{\sc
i}]}&{H~{\sc i}]}&$N({\rm H~I})$&Refs.\\
\hline
Fairall~9~(MS) &\dots&\dots&\ \ $-0.6$&$>-1.2$&\dots&\dots&19.97&{[1,2]}\\
NGC~3783~(MS)  &$>-1.9$&\dots&\ \ $-0.6$&\ \ \ \
$-0.8$&$-1.5$&$<-0.2$&19.90&{[3,4]}\\
Mrk~279~(C)    &$<-1.2$&$-0.5$&\ \ $-0.3$&\ \ \ \
$-0.4$&$-0.8$&$<-0.9$&19.49&{[5,6]}\\
Mrk~290~(C)    &\dots&\dots&\ \ $-1.1$&$>-1.6$&\dots&\dots&19.96&{[5,7]}\\
Mrk~817~(C)    &$<-1.4$&$-0.5$&\ \ $-0.4$&\ \ \ \
$-0.5$&$-0.6$&$<-1.1$&19.48&{[5,6]}\\
Mrk~876~(C)    &\ \ \ \ $-1.4$&\dots&\ \ \dots&\ \ \ \
$-0.2$&$-0.4$&$<-0.9$&19.37&{[5,6,8]}\\
PG 1259+593~(C)&\ \ \ \ $-1.8$&$-0.9$&\ \ $-0.8$&\ \ \ \ $-0.8$&$-1.0$&\ \
$-1.24$&19.92&{[6,9]}\\
3C 351~(C)     &       $<-1.1$&$-0.7$&$<0.1$&\ \ \ \
$-0.4$&$-0.3$&\dots&18.62&{[10]}\\
\hline
ioniz. corr.&{\bf [N/H]}&{\bf [O/H]}&{\bf [S/H]}&{\bf [Si/H]}&{\bf
[Fe/H]}&{\bf [Ar/H]}\\
\cline{2-7}
3C 351 (C)&${\bf <-1.4}$&${\bf-0.8}$&${\bf <-0.3}$&${\bf -0.7}$&${\bf
-0.5}$&\dots&&{[10]}\\
\hline
\end{tabular}\label{hivel_abund}
Key to references: [1] Gibson et al. 2000, [2]  Lu, Savage, \& Sembach
1994, [3]  Lu et al. 1998, [4]  Sembach et al. 2001, [5]  Gibson et al.
2001, [6]  Collins, Shull, \& Giroux 2003, [7]  Wakker et al. 1999, 
[8]  Murphy et al. 2000, [9]  Richter et al. 2001b,
[10] Tripp et al. 2003.  The values of [O~{\sc i}/H~{\sc i}] listed by
Collins et
al. (2003) have been adjusted to a reference solar abundance of 8.69
given by Allende Prieto et al. (2001).
}
\end{table*}

For sightlines that penetrate through well-known high-velocity cloud
complexes called Complex~C and the Magellanic Stream,
Table~\ref{hivel_abund} lists the observed deficiencies of the dominant
ion $X_i$ of element $X$ in H~{\sc i} regions with respect to neutral
hydrogen,
\begin{equation}\label{ionratio}
[X_i/{\rm H~I}]=\log\{N(X_i)/N({\rm H~I})\}-\log(X/{\rm H})_\odot,
\end{equation}
from which one may define a true depletion of the element
\begin{equation}\label{corr}
[X/{\rm H}]=[X_i/{\rm H~I}]+\log\{f({\rm H~I})/f(X_i)\}
\end{equation}
if the ion ratio $f({\rm H~I})/f(X_i)$ can be calculated.  Tripp et al. 
(2003) argue that for the Complex~C gas in front of 3C~351 the
corrections of Equation~\ref{corr} are substantial and are most likely to
arise from collisional ionization.  The last row of
Table~\ref{hivel_abund} shows the element abundances after these
corrections were made.  One can argue that the other four directions
through Complex~C are probably less susceptible to ionization effects,
particularly from photoionization, because their high-velocity gas
components have much larger values of $N$(H~{\sc i}) than that toward 3C~351. 
While this may be true, the fact that [Ar~{\sc i}/H~{\sc i}]~$<$~[O~{\sc
i}/H~{\sc i}] for
Mrk~279, Mrk~817, and PG~1259+593 warns us that ionization corrections for
other elements, while probably small, are not completely
inconsequential.  (Of all the elements, Ar is the most responsive one to
the effects arising from partial photoionization.)

The Complex~C determinations shown in Table~\ref{hivel_abund} indicate
that the generally very mildly depleted species N and O (see
\S\ref{reinvestigation}) are substantially below their corresponding
solar abundances.  The findings for O are particularly secure since
their corrections arising from Equation~\ref{corr} are very small (e.g.,
0.03~dex\footnote{In Table~\protect\ref{hivel_abund} the correction
appears to be 0.1~dex, but this is due to roundoff error.} for 3C~351),
principally because the ionization of O is strongly coupled to that of H
through charge exchange reactions that have large rate constants (see
\S\ref{ionization}).  The fact that N seems even more deficient than O
[even after applying the corrections, which amount to $<0.2$~dex
according to Tripp et al.  (2003)] indicates that the gas in Complex~C
has not only a low metallicity, but it has not undergone multiple
generations of secondary element processing that are characteristic of
the disk of our Galaxy  (Henry, Edmunds, \& K\"oppen 2000).  These two
considerations disfavor the proposition that Complex~C arose from a
recent ejection of gas from the Galactic disk, even after some heavy
dilution from more pristine material in the Galactic halo or Local
Group.  Finally, the fact that [Fe/H] is not substantially less than the
values for other elements indicates that the fraction of elements locked
into grains is much smaller than for gas within the Galactic disk.  This
finding is consistent with an apparent lack of H$_2$ for material arising
from Complex~C in front of PG~1259+593  (Richter et al. 2001a), since H$_2$
is produced primarily on the surfaces of grains.  A question may arise about
a possible enhancement of Fe over $\alpha$-capture elements, but Tripp et al.
caution that currently the errors in both the measurements and correction
factors are too large to make a sound evaluation of this proposition.

The picture for the Magellanic Stream is not as complete as that for
Complex~C.  Nevertheless, the pattern appears to be very similar, except
for evidence that Fe is deficient toward NGC~3783, which in turn
suggests that some dust is present [further support for the presence of
dust in the Magellanic Stream is suggested by the presence of H$_2$  (Richter
et al. 2001a; Sembach et al. 2001)].  The evidence presented in
Table~\ref{hivel_abund} is consistent with the picture that the Magellanic
Stream is material that has been tidally stripped from the Magellanic Clouds.

\section{Application to Observations of Abundances
Elsewhere}\label{application}

The topics discussed in the preceding sections all lead to interesting
insights on the nature of processes that govern the apparent abundances
of elements in the gas phase within and near our own Galaxy.  A new and
exciting frontier is the exploration of abundances in very distant
systems (Prochaska, Howk, \& Wolfe 2003; Prochaska et al. 2003a, b), once
again through the observations of absorption lines against
continuum sources (QSOs), but ones at very large redshifts.  This is
a productive way to understand the chemical state of systems that are
otherwise invisible, owing to their great distances.  The lessons we
have learned from studying the interstellar material in an environment
that we know well (our Galaxy) can be applied elsewhere.  For instance,
we may start with the assumption that the patterns of dust depletion are
the same everywhere, regardless of the overall level of metallicity.  At
the very least, this assumption has been shown to hold true for Galaxy
and the LMC by Vladilo  (2002).  Provided one has measurements of two or
more elements with very different indices $A_X$ and $A_{0,X}$, but with
somewhat similar production origins and responses to ionization, it
should be possible to solve simultaneously the different forms of
Equation~\ref{law} to arrive at a representative $F_*$.  Once this has been
done, the $F_*$ and observations of all of the elements may once again
be substituted into Equation~\ref{law} to derive the absolute abundances. 
These, in turn, lead to our understanding of the chemical histories of
such systems, along with key properties of their internal environments.

The preparation of this paper was supported by NASA contract NAS5-30110.  The
author thanks B.~T.~Draine, J.~X.~Prochaska, T.~M.~Tripp, B.~D.~Savage,
and D.~Welty for helpful comments on early drafts of this paper.

\end{document}